\documentclass[doublecol]{epl2}

\newcommand{\bra}{\left\langle}
\newcommand{\ket}{\right\rangle}

\newcommand{\pder}[2]{\frac{\partial #1}{\partial  #2}}

\newcommand{\var}[2]{\frac{\delta #1}{\delta  #2}}
\newcommand{\bv}[1]{{\boldsymbol #1}}
\newcommand{\e}{{\rm e}}

\newcommand{\phic}{\phi_{\rm cl}}
\newcommand{\phis}{\phi_*}
\newcommand{\tphis}{\tilde \phi_*}
\newcommand{\phiB}{\phi_{\rm B}}
\newcommand{\ep}{\epsilon}

\newcommand{\Zt}{Z_\theta}
\newcommand{\tauw}{\tau_{\rm w}}
\newcommand{\Tc}{T_{\rm c}}

\title{A theory for critically divergent fluctuations of dynamical events at non-ergodic transitions }
\shorttitle{Divergent fluctuations at non-ergodic transitions }

\author{Mami Iwata\inst{1} \and Shin-ichi Sasa\inst{1}}
\shortauthor{F. Author \etal}  

\institute{                    
  \inst{1}
  Department of Pure and Applied Sciences, University of Tokyo, Komaba, Tokyo 153-8902, Japan
}


\abstract{
We theoretically study divergent fluctuations of dynamical
events at non-ergodic transitions. We first focus  on the finding 
that a non-ergodic transition can be described as a saddle 
connection bifurcation of an order parameter for  a time correlation 
function. Then, following the basic idea of Ginzburg-Landau theory for 
critical phenomena, we construct a phenomenological framework with
which we can determine the critical statistical properties at saddle 
connection bifurcation points. Employing this framework, we analyze 
a model by considering the fluctuations of an instanton in space-time 
configurations of the order parameter. We then obtain the exponents 
characterizing the divergences of the length scale, the time scale and 
the amplitude of the fluctuations of the order parameter at the saddle 
connection bifurcation. The results are to be compared with those of 
previous studies of the four-point dynamic susceptibility 
at non-ergodic transitions in glassy systems.
}

\begin{document}

\maketitle


Slow relaxation behavior is observed generally in glassy systems, 
such as super-cooled liquids, dense colloidal suspensions, 
foams, emulsions and granular materials. 
In particular, it has been found that for some systems,
the time correlation function exhibits a plateau over 
a certain time interval. Although the question of whether or 
not the plateau becomes infinitely long at some transition 
temperature has not been definitely answered, there is
experimental evidence that the plateau does become very long 
as a particular temperature is approached. 
This distinctive phenomenon is called a non-ergodic 
transition \cite{Gotze}. 
Recently, in addition to studies of the singular behavior of glassy
systems, there have been investigations of  the growth of the 
characteristic length scale and the enhancement of the four-point 
dynamic susceptibility with the goal of elucidating the nature of 
the cooperative fluctuations of dynamical events 
responsible for the slow relaxation \cite{review}.

As one means for quantifying dynamical events in glassy systems,
the following fluctuating quantity has been considered 
\cite{Laceviv,Dauchot}:
\begin{equation}
q_a(\bv{x},t)=\delta\rho(\bv{x},t)\int d^3 \bv{y}
w_a(\bv{x}-\bv{y})\delta\rho(\bv{y},0),
\label{qa}
\end{equation}
where $\delta\rho$ represents the deviation of the local density from 
its  ensemble average and $w_a(\bv{r})=\exp(-|\bv{r}|^2/(2a^2))$ is 
an overlap function with size $a$. If $q_a(\bv{x},t)$ differs significantly
from $q_a(\bv{x},0)$, it can be concluded that a dynamical event leading 
to the de-correlation of the local density occurs during the time interval 
$[0,t]$. The ensemble average $\bra q_a(\bv{x},t) \ket$ is the 
time-correlation function of the local density field. For the system
with the spatially translational symmetry,  it does not depend on 
$\bv{x}$. Using the quantity $q_a$, fluctuations of dynamical events 
are naturally quantified by  the function
$C_4(\bv{r},t)\equiv \bra q_a(\bv{r},t)q_a(\bv{0},t) \ket-
\bra q_a(\bv{0},t)\ket^2$. Then, it is natural to 
define the amplitude of fluctuations 
as $\chi_4(t)\equiv  \int d^3 \bv{r} C_4(\bv{r},t)$. 
It has been conjectured that $\chi_4(t)$ has a peak at some time
$t=t_* \sim |T-\Tc|^{-\zeta}$, that $\chi_4(t_*)$ diverges as 
$|T-\Tc|^{-\gamma}$, 
and that the correlation length $r_*$ of $C_4(\bv{r},t_*)$ diverges as  
$|T-\Tc|^{-\nu}$, where $\Tc$ is the temperature at which a non-ergodic 
transition occurs \cite{Laceviv}.

There are several theoretical approaches to this phenomenon.  
A reliable one might be mode coupling 
theory, because it has been recognized that the theory can
provide  accurate calculation  for some experimental results
of glassy systems. Recently, within a framework of mode coupling 
theory, the exponents 
of the divergences are calculated \cite{BB,Miya}. 
In contrast, as an alternative approach, one seeks for another
framework that is independent 
of specific properties of glassy systems. For example, 
a space-time thermodynamic formalism has been proposed with
this motivation \cite{garrahan}. In this Letter, we present
a new theory  in the latter approach.  The framework 
we propose is analogous to the Ginzburg-Landau theory of critical 
phenomena \cite{GL}. Within this framework, we calculate the exponents 
of the divergences by applying a singular perturbation method to
a path integral expression. 

\section{Framework:}
Let us begin by recalling that the effective Hamiltonian in 
Ginzburg-Landau theory describes a pitchfork bifurcation in 
the simplest form; that is, the Ginzburg-Landau potential 
possesses a double well form  below a  critical temperature.
This simplicity is responsible for the universality of the 
results of the Ginzburg-Landau theory. With this property
of Ginzburg-Landau theory in mind, the first problem we 
consider here is to  identify the type of the bifurcation 
occurring at  non-ergodic transitions. In fact,  this problem
is solved in Ref. \cite{IwataSasa}: 
The behavior of the time correlation function is determined by an
equation, $\partial_t^2 \phi=f_\ep(\phi,\partial_t\phi)$ 
that exhibits  a saddle connection bifurcation \cite{bifur}.
Here, the order parameter $\phi$ is defined as follows.
First, the equation for the time correlation function
$C(\bv{r},t)(\equiv \bra \delta\rho(\bv{r},t)\delta\rho(\bv{0},0)\ket )$ 
is derived as  $\partial_t^2 C={\cal F}(C)$
from a many-body Langevin equation, 
under the assumption that third-  and fourth-order
cumulants of density fluctuations are ignored in 
$\partial_t^2 C$. We then find that the linear operator 
$\delta {\cal F}/\delta C$ around $C=0$  has a zero eigenvalue
at some temperature $T=T_0$ with decreasing $T$.  Using 
the zero eigenvector $\Phi_{00}(\bv{r})$ that satisfies 
$\delta {\cal F}/\delta C \cdot \Phi_{00}=0 $ at $T=T_0$, 
we define  the order parameter $\phi(t)$ as the amplitude
of $\Phi_{00}(\bv{r}) $ for $C(\bv{r},t)$. See Ref.\cite{IwataSasa}
for the detail. 

Here, we note that the results presented in this work are independent of
the form of $f$, as long as  it exhibits a saddle connection bifurcation 
at $\ep=0$ with decreasing $\ep$, as we find below. For this reason,
instead of employing the complicated form of  $f_\ep(\phi,\partial_t\phi)$
 derived for a specific model of many-body Langevin systems in 
Ref. \cite{IwataSasa},  we assume the simple form 
$f_\epsilon(\phi,\partial_t\phi)=-\partial_\phi U_{\ep}(\phi)$
with $U_{\ep}(\phi)=-\phi^2\left[(\phi - 1)^2+\epsilon\right]$, where
$\ep \ge 0$ \cite{fn:diss}. The existence of the  saddle connection 
bifurcation in this model is easily seen  from the energy conservation 
condition 
$(d\phi/dt)^2/2+U_{\ep}(\phi)=0$ for trajectories of $\phi(t)$ 
that satisfy  a boundary 
condition $\phi(\infty)=0$. This equation 
exhibits the following behavior. 
When $\ep$ is small, $\phi$ first approaches $\phi=1$ 
from $\phi(0)=\phi_0 >1$, and then after a delay, it begins to 
approach the origin, $\phi=0$. As $\ep \to 0$, the time interval
during which $\phi$ remains near $1$ diverges. This type of
bifurcation is termed a saddle connection bifurcation,  
because the two fixed points $\phi=1$  and $\phi=0$, which form
saddles in the two-dimensional phase space $(\phi,\partial_t\phi)$,
are  connected by a trajectory for the case $\ep=0$. There is a
clear qualitative correspondence between $\phi(t)$ 
($0 \le t \le \infty$) and a time correlation function exhibiting 
a non-ergodic transition.

Now, we employ the basic idea of Ginzburg-Landau theory. 
First, in order to describe large-scale spatial variation
of the order parameter $\phi(t)$, 
we replace the order parameter $\phi(t)$ with 
an order parameter field $\phi(\bv{x},t)$ \cite{GL}. 
When  fluctuations of $\phi$ are ignored, 
as the simplest description of  spatially large-scale 
behavior of the order parameter field, 
we assume that $\phi(\bv{x},t)$ obeys the evolution equation 
$\partial_t^2 \phi=f_\ep(\phi) +D \Delta^2 \phi$ under the 
boundary condition $\phi(\bv{x},\infty)=0$,
where  $D \Delta^2 \phi$ represents  the  diffusive coupling 
associated with $\partial_t^2\phi$, and we write 
$f_\epsilon(\phi,\partial_t\phi)$ as $f_\epsilon(\phi)$,
for simplicity.
Here, $\Delta^2=(\bv{\nabla} \cdot \bv{\nabla})^2$.
Note also that
periodic boundary conditions are first assumed for the system with a 
size $L$, and then the limit $L \to \infty$ is taken. 
Next, we take account of the fluctuation effects 
of $\phi(\bv{x},t)$ by considering 
 space-time configurations of $\phi(\bv{x},t)$ with 
($\bv{x} \in {\bf R}^3$ and $ 0 \le  t \le \infty$), which we denote by
 $[\phi]$. 
Then, let $P([\phi])$ be the probability measure for $[\phi]$.
Until now, we have not derived the accurate form of $P([\phi])$ 
on the basis of microscopic models. Nevertheless, relying
on a phenomenological argument, we can reasonably assume 
a form of $P([\phi])$ in the following manner. 

First, we divide  time $t$ 
as $t_k =k \Delta t$, with $k=1,2 \cdots$. Denoting a collection 
of space configurations of $(\phi(\bv{x},t), \partial_t \phi(\bv{x},t))$
at an instant time $t$ as $u(t)$, we assume that there is a time interval 
$\Delta t$ such that $\tau_{\rm m} \ll \Delta t \ll \tau_{\rm M}$, 
where $\tau_{\rm m}$ is the longest time scale of fast variables
eliminated in deriving the order parameter equation from  microscopic
models, and $\tau_{\rm M}$ is the characteristic time scale 
of the time evolution of $u$. We then consider the transition probability
$ T(u  \to  u' ) $ during a time interval $\Delta t$.
Now, as in the 
case of the Onsager-Machlup theory 
for stochastic processes describing the time evolution of thermodynamic variables 
\cite{OM}, the assumption mentioned above suggests 
the following two properties: (i) $T(u  \to  u' )$  can define a Markov 
process and (ii) $T(u \to u')$ possesses the large deviation 
property with respect to time. 
From the first property (i), we can express $P([\phi])$  as 
\begin{equation}
P([\phi]) \simeq
P_0(u_0)
T(u_0\to u_1)
T(u_1\to u_2)
\cdots,
\label{P:T}
\end{equation}
in the discretized description, where $u_k=u(k\Delta t)$ and 
$P_0(u_0)$ is the stationary distribution function of $u$ at an instant 
time. Further, from the second property (ii), 
$T$ can be expressed as
$T(u \to u') \simeq \exp(-\Delta t I( u'|u))$
with a large deviation functional $I(u'|u)$. 
Here, $I(u'|u)=0$ should provide the discretized form of 
the deterministic equation 
$\partial_t^2 \phi=f_\epsilon(\phi)+D\Delta^2 \phi$. 
Thus, approximating $I( u'|u)$ with the second order polynomial 
of $u'$, we assume 
\begin{eqnarray}
&&I( u'|u)=
\frac{1}{2B}\int d^3\bv{x}\nonumber\\
&&[(\partial_t \phi)'-\partial_t \phi -\Delta t 
(f_\epsilon(\phi)+D\Delta^2 \phi)]^2/(\Delta t)^2,
\label{LD}
\end{eqnarray}
where $B$ is a constant. From (\ref{P:T}) and (\ref{LD}), 
taking the limit $\Delta t/ \tau_M \to 0$,
we can express 
$P([\phi])$ as 
\begin{equation}
P([\phi])=\frac{1}{Z} \exp \left[-\frac{1}{B}F([\phi]) \right ]
\label{st-measure}
\end{equation}
with
\begin{equation}
F([\phi])= \frac{1}{2}\int_0^\infty dt \int d^3\bv{x} 
(\partial_t^2 \phi-f_\epsilon(\phi) -D \Delta^2 \phi)^2 
\label{action}
\end{equation}
under the boundary condition $\phi(\bv{x},\infty)=0$.
Here, instead of considering $P_0(u_0)$,  we fix the 
initial condition  $\phi(\bv{x},0)$ as $\phi_0$ that is 
determined from the equal-time correlation function.  
Note that the so-called Jacobian term is not written 
in (\ref{action}), because it is independent of $[\phi]$
in the present problem \cite{Hoch}.  

Although the model (\ref{st-measure}) with (\ref{action}) 
is not justified by the
analysis of microscopic models, we expect that this model
can capture essential aspects of fluctuations near the
saddle connection bifurcation. With this expectation,
we calculate the quantity 
$\bra \phi(\bv{x},t) \ket$ and 
\begin{equation}
C_{\phi}(\bv{x},t)\equiv 
\bra \phi(\bv{x},t) \phi(\bv{0},t) \ket-
\bra \phi(\bv{x},t)\ket \bra  \phi(\bv{0},t) \ket
\end{equation}
employing  (\ref{st-measure}), and then from this we compute
$\chi_\phi(t)\equiv \int d^3\bv{x} C_{\phi}(\bv{x},t)$.
Because  $\bra \phi(\bv{x},t) \ket $ determines the 
behavior of $\bra q_a(\bv{x},t)\ket$,
we believe  that $\phi(\bv{x},t)$  characterizes  dynamical 
events responsible for slow relaxation. We thus conjecture that 
$\chi_\phi(t)$ has a peak at some $t=t_*$ and that the system 
displays divergences of the forms
$t_* \simeq \ep^{-\zeta}$, $\chi_\phi(t_*) \simeq \ep^{-\gamma}$, 
and $r_* \simeq \ep^{-\nu}$, where $r_*$ is the correlation length 
defined from $C_{\phi}(\bv{x},t_*)$.
In the following, we demonstrate that indeed such divergences do 
occur and that we have the values $\zeta=2$, $\gamma=3/4$ and $\nu=1/4$.


\section{Analysis:}

The probability distribution (\ref{st-measure}) can be realized as 
the stationary distribution of the four dimensional field $\phi(\bv{x},t)$ 
that obeys a fictitious stochastic process. Introducing a fictitious 
time $s$, we can construct such a stochastic model as
\begin{equation}
\partial_s \phi(\bv{x},t;s)=-\var{F}{\phi(\bv{x},t;s)}
+\xi(\bv{x},t;s),
\label{eq:p}
\end{equation}
where 
\begin{equation}
\bra \xi(\bv{x},t;s) \xi(\bv{x'}, t';s') \ket
= 2B \delta^3(\bv{x}-\bv{x'})\delta(t-t')\delta(s-s').
\label{noise}
\end{equation}
It is easily checked that the $s$-stationary distribution 
function of $\phi(\bv{x},t)$ for this model is exactly 
expressed by (\ref{st-measure}).
Then, using  the operator
$ \hat L_\ep(\phi) \equiv \partial_t^2- \partial_{\phi}f_\ep(\phi)$, 
we write (\ref{eq:p}) explicitly as 
\begin{equation}
\partial_s \phi=-(\hat L_\ep -D \Delta^2)
(\partial_t^2 \phi-f_\ep(\phi)-D\Delta^2 \phi)
+\xi.
\label{eq:p2}
\end{equation}

Now, 
let $\phic(t;\ep)$ be the solution of $\partial_t^2 \phi=f_\ep(\phi)$
with the boundary conditions $\phic(0;\ep)=\phi_0$ and $\phic(\infty;\ep)=0$.
Then, if $B=0$, all solutions of (\ref{eq:p2}) approach  this
solution  in the limit $s \to \infty$. Furthermore, let
$\phis(t)$ and $\phiB(t)$ be  special solutions of 
$\partial_t^2 \phi=f_0(\phi)$ satisfying the conditions 
$\phis(-\infty)=1$, $\phis(\infty)=0$,  $\phiB(0)=\phi_0$, and 
$\phiB(\infty)=1$. Such solutions  exist generally at saddle connection 
bifurcation points. In the model under investigation, 
we have $\phis(t)=(1-\tanh(t/\sqrt{2}))/2$,
which  represents a kink solution in the 
$t$ direction, a kind of instanton \cite{instanton}. Note also that 
$\phiB(t) \simeq \delta \e^{-\lambda t} +1$ for $\lambda t \gg 1$ with
the constants $\lambda$ and $\delta$  determined by the equation
$\partial_t^2 \phi=f_0(\phi)$. Thus, for small  $\ep$,
the solution $\phic(t;\ep)$ is close to 
$\phis(t-\theta_0)+\phiB(t)-1$ with a constant $\theta_0$.
Based on these preliminary considerations, we find that, 
when $\epsilon$ and $B$ are small, it is convenient to express 
solutions of (\ref{eq:p2}) as 
\begin{equation}
\phi(\bv{x},t;s)=\phis(t-\theta(\bv{x},s))+\phiB(t)-1+\varphi(\bv{x},t;s),
\label{set}
\end{equation}
where $\theta(\bv{x},s)$ represents the kink position,
which depends on $(\bv{x},s)$, and $|\varphi|$ is 
assumed to be small 
when $\theta \lambda \gg 1 $ and 
$\epsilon \ll 1 $. We then find  that perturbations of the 
instanton  decay exponentially,  except for the Goldstone mode, 
corresponding to the translation of the instanton in the $t$ direction
\cite{kink_space}.
It is known that in such a situation 
a singular perturbation method is useful to derive 
an evolution equation for the kink position $\theta(\bv{x},s)$ under 
the influence of perturbations taking the forms of  finite epsilon
effects, $\phiB(t)-1$ (see (\ref{set})), and noise \cite{kink}. 

This perturbative calculation is carried out by 
substituting (\ref{set}) into (\ref{eq:p2}) and  extracting 
the important terms under the conditions $\epsilon \ll 1 $, 
$B \ll 1$, $\theta \lambda\gg 1$ and that  spatial 
variation of $\theta$ is small. In order to make these assumptions
explicit, we replace $\phiB-1$, $\ep$, $\nabla$, and $\xi$  with 
$\mu (\phiB-1)$, $\mu \ep$, $\mu^{1/4}\nabla$ and $\mu^2 \xi$ and
consider the  expansions $\partial_s\theta =
\mu\Omega_1+\mu^2 \Omega_2+\cdots$  and $\varphi =
\mu\varphi_1+\mu^2 \varphi_2+\cdots$,
where $\mu$ is a small parameter,
and   $\Omega_i$ and $\varphi_i$ are functions of the field
$\theta(\bv{x},s)$. These replacements are determined in an 
essentially unique manner by the requirement
that a systematic perturbative expansion can be carried out.
Then, collecting the terms proportional to $\mu$, we obtain
\begin{equation}
\Omega_1\partial_t {\phis}-\hat L_{0*}^2 (\phiB-1)=  
\hat L_{0*}[\hat L_{0*} \varphi_1-\ep f_{\rm p}(\phis)
-D\Delta^2 \phis],
\label{sub:1}
\end{equation}
where $\hat L_{0*}=\hat L_{0}(\phis)$ and 
$f_{\rm p}(\phi)=df_{\ep}(\phi)/d\ep|_{\ep=0}$.
Because $\hat L_0 \partial_t {\phis}=0$, the solvability 
condition for (\ref{sub:1}) leads to  $\Omega_1=0$ \cite{fn:kink}.
Employing this condition, 
we obtain $\varphi_1$. In a similar manner, the solvability 
condition for the equation consisting of terms proportional to 
$\mu^2$ yields $\Omega_2$, from which we derive 
\begin{equation}
c_1 \pder{\theta}{s}
= \ep c_2 e^{-\lambda \theta}-\ep^2 c_3-\ep c_4 D \Delta^2 \theta
- c_1D^2\Delta^4 \theta + \Xi,
\label{kd:evol}
\end{equation}
where we have ignored  terms non-linear in the gradient of $\theta$,
such as $(\Delta \theta)^2$ and $\e^{-\lambda \theta} \Delta \theta$, 
and $\Xi$ satisfies 
\begin{equation}
\bra \Xi(\bv{x},s) \Xi(\bv{x'},s') \ket=2 c_1 B \delta^3(\bv{x}-\bv{x'})
\delta(s-s').
\label{kd:noise}
\end{equation}
Note that the coefficients $c_i$ are positive constants 
independent of $\ep$ and can be 
calculated from (\ref{eq:p2}).  The first term on the right-hand side 
of (\ref{kd:evol}) represents  the interaction with the exponential tail
of the solution $\phiB(t)$. The second term in (\ref{kd:evol}) 
determines the velocity of the steady propagation of the kink in a system 
without a boundary in the case $\ep \not =0$.  

Now, we investigate the $s$-stationary statistical properties of $\theta$
determined by (\ref{kd:evol}) with (\ref{kd:noise}). We first consider 
the case $D=0$, in which  the field $\theta$ behaves independently at each 
$\bv{x}$. Thus, according to the central limit theorem,
the spatial average of $\theta$ over a domain $\Omega 
\subset {\bf R}^3$ obeys a Gaussian distribution  
with average $\theta_*$ and  variance $\chi_\theta/|\Omega|$
if the volume of the domain $|\Omega|$ is sufficiently large.
In this case, the leading terms of $\theta_*$ and 
$\chi_\theta$ in the limit $\ep \to 0$  can be calculated 
in an elementary manner and we obtain 
\begin{eqnarray}
\theta_* &=&  \frac{B}{c_3\ep^2}, \label{thetas}\\
\chi_\theta &=&  \frac{B^2}{c_3^2\ep^4} \label{chis}.
\end{eqnarray}
Next, we study the case in which  $D$ is a small positive number. 
Specifically,  defining
\begin{equation}
\tilde  \theta_1(\bv{k},s) \equiv  \int d^3 \bv{x}e^{i \bv{k}\bv{x}} 
( \theta(\bv{x},s)-\theta_*),
\end{equation}
we consider the $s$-stationary distribution function 
\begin{equation}
P_\theta(\{\tilde \theta_1 \})= \frac{1}{\tilde \Zt}
\e^{-\frac{\tilde V(\{\tilde \theta_1 \})}{B}}.
\label{st:k}
\end{equation}
Although it is difficult to derive the exact form of 
$\tilde V$, we conjecture that $\tilde \theta(\bv{k})$ with 
small $|\bv{k}|$ obeys a Gaussian distribution. Then, 
the fluctuation intensity of $\tilde \theta(\bv{k}=\bv{0})$ 
should be equal to $\chi_\theta$. 
Also, considering (\ref{kd:evol}),  we find that the length scale 
of $\theta$ 
is simply determined by the balance among the second, third and 
fourth terms in the right-hand side when we focus on the regime 
$\theta \lambda \gg 1$. This leads to the scaling relation 
$|\bv{k}|^4 \sim \ep $. 
On the basis
of these  two considerations, we assume the simple form
\begin{equation}
\tilde V(\{\tilde \theta_1\})=\int  \frac{d^3\bv{k}}{(2\pi)^3}
\left[\frac{B}{2\chi_\theta} |\tilde \theta_1(\bv{k})|^2
+\frac{D_4}{2} \ep^3 k^4 |\tilde \theta_1(\bv{k})|^2\right],
\label{assum:k}
\end{equation}
where $D_4$ is a constant. Strictly speaking, (\ref{assum:k}) is
valid only for small $|\bv{k}|$, but we believe that 
it provides a good description of the large scale behavior in
which we are interested. We therefore use this expression 
in order to evaluate $\bra \phi(\bv{x},t) \ket$
and $\bra \phi(\bv{x},t)\phi(\bv{x'},t') \ket$ through 
the relation (\ref{set}).

We calculate these quantities by first expressing $\phis(t)$ as
\begin{equation}
\phis(t)=\int \frac{dz}{2 \pi } \tphis \e^{izt}.
\label{zexp}
\end{equation}
Then, using $\bra \phi(\bv{x},t) \ket \simeq   \bra 
\phis(t-\theta(\bv{x})) \ket$,  we  write
\begin{equation}
\bra \phi(\bv{x},t) \ket
\simeq  \int \frac{dz}{2 \pi }
\tphis(z)
\e^{iz(t-\theta_*)} 
\bra \e^{-iz \theta_1(\bv{x}) }\ket .
\label{form1:1}
\end{equation}
From the Gaussian nature of the distribution function 
(\ref{st:k}) with (\ref{assum:k}), we obtain
\begin{equation}
\bra \e^{-iz \theta_1(\bv{x})} \ket 
= \e^{ -\tauw^2 z^2/2},
\label{form1:2}
\end{equation}
with $\tauw =c_5  \ep^{-13/8}$, where $c_5$ is independent of $\epsilon$.
Thus, $\bra \phi(\bv{x},t) \ket$ can be expressed as a
scaling form $g((t-\theta_*)/\tau_{\rm w})$. 
Also, it is seen  that 
$\bra \phi(\bv{x},t) \ket$  possesses a kink in the $t$ direction
whose  average position and width are $\theta_* \simeq O(\ep^{-2})$ 
and $\tauw \simeq O(\ep^{-13/8})$, respectively. 

Next, we calculate $\bra \phi(\bv{x},t)\phi(\bv{x'},t) \ket$ 
by approximating it with $
\bra \phis(t-\theta(\bv{x}))\phis(t-\theta(\bv{x'})) \ket$.
Then,  using (\ref{zexp}), we write 
\begin{eqnarray}
\bra \phi(\bv{x},t)\phi(\bv{x'},t) \ket
&& \simeq   \int \frac{dz}{2 \pi }\int \frac{dz'}{2 \pi }
\tphis(z) \tphis(z') \e^{izt+iz't} 
\nonumber\\
&&
\bra \exp\left [ i \int d^3 \bv{x''} 
J^{\rm s}(\bv{x''}) 
\theta(\bv{x''}) \right] \ket ,
\label{formula}
\end{eqnarray}
with
$
J^{\rm s}(\bv{x''})
= -z \delta^3(\bv{x''}-\bv{x})
  -z' \delta^3(\bv{x''}-\bv{x'}).
$
Performing the Gaussian integration of $\theta$, we  obtain
\begin{eqnarray}
&&
\bra \phi(\bv{x},t)\phi(\bv{x'},t) \ket
\simeq 
\int \frac{dz}{2 \pi }\int \frac{dz'}{2 \pi }
\tphis(z) \tphis(z') \nonumber \\
& & 
\e^{iz(t-\theta_*)+iz'(t-\theta_*)} 
\bra \e^{-iz \theta_1(\bv{x}) }\ket
\bra \e^{-iz' \theta_1(\bv{x'})} \ket 
\nonumber \\
& & 
\exp\left[ -zz'\tauw^2G(c_6 |\bv{x}-\bv{x'}| \ep^{1/4} )\right],
\label{23}
\end{eqnarray}
where $c_6$ is independent of $\epsilon$ and $G(r)= \e^{-r}\sin(r)/r$.  
Specifically, in  the long distance regime, for which we have
$|G(c_6 |\bv{x}-\bv{x'}| \ep^{1/4}) |\ll 1$,  
we can derive the approximate expression
\begin{equation}
C_{\phi}(\bv{x},t) 
\simeq
\tauw^2\left[ \partial_t \bra \phi(\bv{0},t)\ket \right]^2
G(c_6 |\bv{x}| \ep^{1/4}).
\label{asymp}
\end{equation}
From this expression (\ref{23}), we immediately find that the correlation length 
is proportional to  $\ep^{-1/4}$. Thus, we obtain $\nu=1/4$. 
Next, noting that the singular part of $\chi_\phi(t)$ originates from 
the long-distance part of $C_{\phi}(\bv{x},t)$, we can use the asymptotic 
form (\ref{asymp}) in order to evaluate the exponents $\zeta$ and $\gamma$. 
First, it is found  that $\chi_\phi(t)$ has a peak at $t=\theta_*$,
and thus, from (\ref{thetas}), we obtain $\zeta=2$. Furthermore, 
from the relation $\tau_{\rm w} \partial_t g((t-\theta_*)/\tau_{\rm w}) 
\sim O(\ep^{0})$,
we obtain $\chi_\phi(\theta_*) \simeq \ep^{-3/4}$, and hence $\gamma=3/4$.  
Finally, if the asymptotic formula (\ref{asymp}) is valid
 on the  correlation length scale, it
 suggests that the critical  exponent $\eta$ 
 in the standard notation of  critical phenomena 
 is equal to $-1$.


\section{Conclusion:}

We have presented a phenomenological framework for determining
the statistical properties of dynamical events at non-ergodic transitions.  
Within this framework, we
have obtained the results $\zeta=2$, $\gamma=3/4$ and $\nu=1/4$, 
which are to be compared with the values obtained in numerical experiments 
($\gamma/\zeta\simeq 0.4$ and $\nu/\zeta\simeq 0.22$)  \cite{WBG} 
and the values calculated using the mode coupling theory 
($\zeta\simeq 2.3$ for a Lennard-Jones fluid, $\gamma=1$ \cite{BB,Miya}, 
and
$\nu=1/2$ \cite{BB} and $\nu=1/4$ \cite{Miya}.) 
\cite{BBBMR}.
It might be rather surprising that our theoretical values 
are close to the results of the mode coupling theory despite 
the great difference between the two frameworks. However, 
because both theories are still at the mean field level in 
the sense that critical fluctuations are not treated precisely, 
it is not time to compare the results in detail. Similarly,
with regarding the discrepancy with the results by numerical 
simulations, we wish to be cautious of a hasty judgment. 
Rather, the 
important result specific to our theory is that 
the divergence of the  fluctuation is caused by 
the Goldstone mode for the instanton. We expect that this mechanism
should be taken into account by combining a singular perturbation method
with a path integral approach for the analysis of  microscopic models.


\acknowledgments

The authors thank K. Kawasaki, K. Miyazaki and H. Tasaki 
for useful comments on the manuscript.
This work was supported by a grant from the Ministry of Education, 
Science, Sports and Culture of Japan (No. 16540337).


\begin{thebibliography}{10}

\bibitem{Gotze}
  \Name{G\"otze W.}
  \Book{Liquids, Freezing and Glass Transition}
  \Editor{Levesque D. et al}
  \Publ{Elsevier, New York}
  \Year{1991}.

\bibitem{review}
See [
  \Name{Berthier L. et al}
  \REVIEW{Science}{310}{2005}{1797}.] as the latest 
experimental report. See also Introduction in \cite{Laceviv}
for a review of studies. 

\bibitem{Dauchot}
  \Name{Dauchot O., Marty G. \and Biroli G.}
  \REVIEW{Phys. Rev. Lett.}{95}{2005}{265701};
  \Name{Donati C. et al}
  \REVIEW{J. of Non-Cryst. Solids}{307}{2001}{215}.

\bibitem{Laceviv}
  \Name{La\v{c}evi\'c N. et al}
  \REVIEW{J. Chem. Phys.}{119}{2003}{7372}.

\bibitem{BB}
  \Name{Biroli G. \and Bouchaud J. -P.}
  \REVIEW{Europhys. Lett.}{67}{2004}{21}.

\bibitem{Miya}
  \Name{Biroli G., Bouchaud J. -P., Miyazaki K. \and Reichman D. R.}
  \REVIEW{Phys. Rev. Lett.}{97}{2006}{195701}.

\bibitem{garrahan}
  \Name{Merolle M., Garraham J. P., \and Chandler D.}
  \REVIEW{Proc. Natl. Acad. Sci. USA}{102}{2005}{10837}.


\bibitem{GL}
  \Name{Goldenfeld N.}
  \Book{Lectures on Phase Transitions and the Renormalization Group}
  \Publ{Addison-Wesley, New York}
  \Year{1992}.

\bibitem{IwataSasa}
  \Name{Iwata M. and Sasa S.}
  \REVIEW{J. Stat. Mech.}{}{2006}{L10003}.

\bibitem{bifur}
  \Name{Guckenheimer J. \and Holmes P.}
  \Book{Nonlinear Oscillations, Dynamical Systems and Bifurcations of Vector Fields}
  \Publ{Springer-Verlag, New York}
  \Year{1983}.


\bibitem{fn:diss}
$f_\ep(\phi,\partial_t\phi)$ is not restricted to  Hamiltonian
forms. Indeed, $f_\ep(\phi,\partial_t\phi)$ obtained 
in Ref. \cite{IwataSasa} includes a non-Hamiltonian term. 



\bibitem{OM}
  \Name{Onsager L. \and Machlup S.}
  \REVIEW{Phys. Rev.}{91}{1953}{1505}.


\bibitem{Hoch}
  \Name{Hochberg D., Molina-Paris C., Perez-Mercader J. \and Visser M.}
  \REVIEW{Phys. Rev.E}{60}{1999}{6343}.

\bibitem{instanton}
  \Name{Rajaraman R.}
  \Book{Solitons and Instantons}
  \Publ{North-Holland, Amsterdam}
  \Year{1987}.


\bibitem{kink_space}
Because kink profiles in the space 
directions disappear in the large $s$ regime, 
the spatially homogeneous solution is chosen 
as an unperturbative state.




\bibitem{kink}
  \Name{Kawasaki K. \and Ohta T.}
  \REVIEW{Physica A}{116}{1982}{573};
  \Name{Ei S. \and Ohta T.}
  \REVIEW{Phys. Rev. E}{50}{1994}{4672}.

\bibitem{fn:kink} The term $\hat L_{0*}^2  (\phiB-1)$ apparently
yields a contribution to $\Omega_1$ of the form $\sim \e^{-\lambda \theta}$,
as observed  in Ref. \cite{kink}, but the coefficient of this contribution 
is zero in the present case. 

\bibitem{WBG}
  \Name{Whitelam S., Berthier L., \and Garraham J. P.}
  \REVIEW{Phys. Rev. Lett.}{92}{2004}{185705}.
  Note, however, that different values
($\zeta\simeq 2.0$, $\gamma\simeq 1.7$ and $\nu\simeq 0.8$) are 
also reported in Ref. \cite{Laceviv}.

\bibitem{BBBMR}
According to the recent paper [
\Name{Berthier L. et al}
cond-mat/0609656 preprint
{2006}.], the value of the exponent 
$\gamma$ is influenced by the presence of the
conservation law.

\end{thebibliography}
\end{document}